# ЭКСПЕРИМЕНТАЛЬНАЯ ОЦЕНКА ТОЧНОСТИ ПРЕДСТАВЛЕНИЯ ПЛАВАЮЩИХ ЧИСЕЛ И ВРЕМЕНИ ВЫПОЛНЕНИЯ БЛОЧНОГО МЕТОДА КОРКИНА-ЗОЛОТАРЕВА ДЛЯ ПРИВЕДЕНИЯ БАЗИСА ЦЕЛОЧИСЛЕННЫХ РЕШЕТОК СЛОЖНЫХ ПО ГОЛЬДШТЕЙНУ-МАЙЕРУ

**В. С. Усатюк**
**Научный руководитель: д.ф.-м.н., проф. О.В. Кузьмин**
**Братский государственный университет,**
**г. Братск, L@Lcrypto.com**

Целями работы является профилировка с целью увеличения скорости работы и подтверждение достоверности результатов полученных комплексом программ LRT [1] осуществляющим приведение базиса целочисленных решеток блочным методом Коркина-Золотарев, BKZ [2].

В работе на основе приведения базиса целочисленных решеток с целью получения базиса, кратчайший вектор в котором отклоняется $\leq 6,1\%$ от верхней оценки Минковского, решаются три задачи:

1. определение необходимой точности представления чисел с плавающей запятой;

2. определение суммарного времени приведения до 93-мерного базиса целочисленных решеток включительно;

3. соотношение временных затрат на ключевые части BKZ, алгоритмы: ортогонализации и поиска кратчайшего вектора стратегией Поста [3].

Для подтверждения результатов использовались целочисленные решетки из международного конкурса[4] сложные по Гольдштейну-Майеру[5], самые сложные в классе случайных целочисленных решеток. Удалось повторить результаты конкурсантов до 103 мерных решеток, а так же улучшить ранее полученные результаты для размерностей 58, 62, 69, 71, 74, 81 занявшие 84, 77, 68, 65, 60 и 49-места (по состоянию на 10.09.12) на конкурсе алгоритмов соответственно, [4].

Ряд необходимых положений теории решеток представлен в статье «Обзор методов приведения базиса решеток и некоторых их приложений» находящейся в этом сборнике.

**Определение 1:** Решетка является целочисленной, если скалярное произведение любой пары векторов принадлежащих ей является целым числом: $\forall b_i, b_j \in L : \langle b_i, b_j \rangle \in Z, i \neq j$. Иными словами, такая решетка является дискретной абелевой подгруппой, заданной на множестве $Z^n$.

Задача приближенного поиска кратчайшего вектора ($\gamma$-approximate shortest vector problem, SVP$^p_\gamma(n)$): Пусть дана $m$-мерная решетка $L$, ранга $n$ и вещественный $\gamma > 0$. Найти ненулевой вектор в $\gamma$ – раз больший кратчайшего вектора $\bar{b} \in LZ^n \setminus \{0\}: \left\|\bar{b}\right\|_p \leq \gamma \cdot \lambda_1^p(L)$. Для решения $SVP^p_\gamma(n)$ с экспоненциальной точностью $\gamma = 2^{\frac{n-1}{2}}$, достаточно привести базис к приведенному по Ловасу базису, применив полиномиальный по временной сложности алгоритм Ленстра-Ленстра-Ловаса(LLL) [6]. Для решения $SVP^p_\gamma(n)$ с точностью $\gamma \in \left[2^{\frac{n-1}{2}} - \varepsilon, 1\right]$, достаточно привести подмножество базиса решетки, состоящее из $\beta \leq n$-векторов, к базису Коркина-Золотарева, применив BKZ, чья временная сложность зависит от размера блока, изменяясь от полиномиальной до экспоненциальной [2], однако растет значительно медленнее алгоритма Финке-Поста, полного перебора в сфере [3].

В соответсвии верхней оценкой константы Эрмита Блихфельдта [7] и первой теоремой Минковского: первый соответствующий минимум (кратчайший вектор) любой $n$-мерной решетки ограничен сверху: $\lambda_1^p(L) \leq \frac{2}{\sqrt{\pi}} \Gamma(1+\frac{n}{2})^{\frac{1}{n}} \det(L)^{\frac{1}{n}}$. Нами решалась задача:

$$\text{SVP}^2_{1.061}(n) \leq 1.061 \frac{2}{\sqrt{\pi}} \Gamma(1+\frac{n}{2})^{\frac{1}{n}} \det(L)^{\frac{1}{n}}.$$

LLL-алгоритм приведения базиса решетки [7]:
Вход: Базис решетки $L(B), L \subset Z^n$ и $\delta \in (0.25, 1]$, $\Delta$.
Выход: Приведённый LLL-базис.

1. $QR$-разложение базиса решетки.

2. Приведение векторов базиса по длине: $|r_{i,j}| \leq \frac{1}{2}|r_{i,i}|, 1 \leq i \leq j \leq n$

3. Перестановка векторов $r_{i,i}$, $r_{i+1,i+1}$, и соответствующих им $b_i$ и $b_{i+1}$, возврат к шагу 1, если: $r_{i,i}^2 + r_{i-1,i}^2 \geq \delta r_{i-1,i-1}^2, 1 < i \leq n, \delta \in (0.25, 1]$.

BKZ-алгоритм приведения базиса решетки [2]:
Вход: Базис решетки $L(B), L \subset Z^n$ и $\beta \leq n$ ( $L(B')$ образованный $\beta$-векторами из $L(B)$ ), $\Delta$. Выход: Приведённый BKZ-базис решетки.

1. Применение LLL-алгоритма для предварительного приведения

базиса решетки;

2. Выполнять в цикле $k = 1\ldots\quad 1$:

Методом Финке-Поста [3] решаем $\text{SVP}^2_\gamma(n)$-задачу для квадратных подматриц размера $\beta$ образованных верхнетреугольной матрицей $R$ в QR-разложении исходного базиса, $L'(R'): \overline{r}_i^{\,'} = \lambda_1^p(L[b_j^{(i)}]_{j \geq i}), i \leq n.$.

-если $\delta \cdot \|r_{i,i}\| > b_j^{(i)}$: вставка найденного вектора в начало текущего блока и удаление линейной зависимости векторов в нем;
-иначе, LLL-приведение следующего блока векторов.

3. Вывод базиса, если после нескольких проходов по всему базису его длина не меняется.

В приложение LRT были реализованы LLL и BKZ алгоритмы приведения базиса решеток. Для *QR*-разложения применяется блочный метод Хаусхолдера[9], эффективный при использовании длинной арифметики [8], реализованный при помощи библиотеки AMD Core Math Library 4.4.0. Приложение написано на языке Си с применением глубокой оптимизации в среде MVS 2010 Professional. Ввод-вывод данных осуществляется на основе работы с текстовыми файлами базиса решетки в формате библиотеки NTL. В приложении доступно изменение параметров: $\Delta$ точности промежуточных результатов, $\delta$ для LLL и $\beta$ для BKZ. Для реализации произвольной точности вычислений применяется библиотека MIRACL переданная автору статьи Майклом Скоттом для не коммерческого применения в 2006 г.. Применение этой библиотеки позволяет изменять $\Delta$ - точность промежуточных результатов от однократной (float), двукратной (double), четырехкратной и до $1{,}7 \cdot 10^6$ десятичных знаков. С целью повышения устойчивости работы приложения при высокой точности вычислений, а так же большой размерности приводимых целочисленных решеток был реализован собственный планировщик памяти, уменьшающий число обращений к менеджеру памяти ОС. Повышение точности представления чисел с плавающей запятой требуется именно на этапе *QR*-разложения, для последующих этапов достаточной является двойная точность. Последнее обстоятельство не позволяет получить заметный выигрыш в производительности за счет применения видеокарт для *QR*-разложения. Время перебора точек в сфере методом Финке-Поста с ростом размерности, растет быстрее *QR*-разложения. Дальнейшая работа направлена на реализацию метода Финке-Поста на видеокартах с поддержкой стандарта OpenCL, так как особенность работы алгоритма позволяет предположить возможность ускорения этого метода.

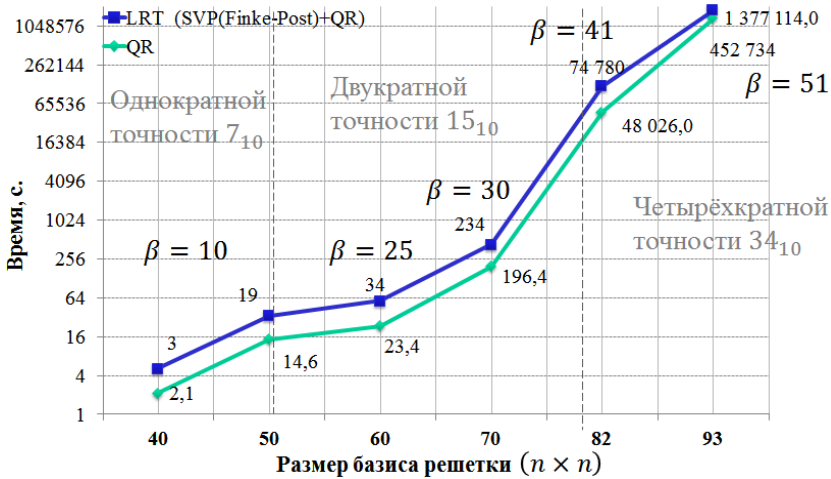

Рис. 1 Зависимость времени приведения базиса целочисленной решетки от ее размерности.